\documentclass[conference]{IEEEtran}
\IEEEoverridecommandlockouts
\usepackage{cite}
\usepackage{pgf}
\usepackage{bm}
\usepackage{amsmath,amssymb,amsfonts}
\usepackage{algorithmic}
\usepackage{graphicx}
\usepackage{textcomp}
\usepackage{xcolor}
\usepackage{comment}
\def\BibTeX{{\rm B\kern-.05em{\sc i\kern-.025em b}\kern-.08em
    T\kern-.1667em\lower.7ex\hbox{E}\kern-.125emX}}

\DeclareMathOperator*{\argmax}{arg\,max}

\begin{document}

\title{An Information Theoretic Analysis of Ghost Modulation}
\author{
\IEEEauthorblockN{Daniel Harman, Ashton Palacios, Philip Lundrigan, Willie K. Harrison}
\IEEEauthorblockA{Department of Electrical and Computer Engineering\\
Brigham Young University\\
Provo, UT, USA\\
\{jdharman, apal6981, lundrigan, willie.harrison\}@byu.edu}
\thanks{This work was partially funded by the US National Science Foundation: Grant Award Numbers \#2135732, \#2153317, and \#2030165.}
}
\maketitle

\begin{abstract}
Side channels have become an essential component of many modern information-theoretic schemes. The emerging field of cross technology communications (CTC) provides practical methods for creating intentional side channels between existing communications technologies. This paper describes a theoretical foundation for one such, recently proposed, CTC scheme: Ghost Modulation (GM). Designed to modulate a low-data-rate message atop an existing network stream, GM is particularly suited for transmitting identification or covert information. The implementation only requires firmware updates to existing hardware, making it a cost-effective solution. However, GM provides an interesting technical challenge due to a highly asymmetric binary crossover erasure channel (BCEC) that results from packet drops and network delays. In this work, we provide a mathematical description of the signal and channel models for GM. A heuristic decision rule based on maximum-likelihood principles for simplified channel models is proposed. We describe an algorithm for GM packet acquisition and timing synchronization, supported by simulation results. Several well known short block codes are applied, and bit error rate (BER) results are presented.

\end{abstract}



\section{Introduction}\label{sec:Introduction}

Side channels have become an important topic within information theory in recent years because of their implications on security \cite{SIDE_CHANNEL_SECURITY} and covertness \cite{SIDE_CHANNEL_HIDING_1, SIDE_CHANNEL_HIDING_2}. They will likely continue to be an important topic moving forward into the era of post-quantum cryptography, with concerns over digital privacy. Most of the existing literature has centered on mitigating information leakage from unintended side channels to preserve security and privacy. This is not always the case, however, as intended side channels can be used for covert communications.

Cross technology communications (CTC) is an emerging field in communications and networking theory that enables devices not explicitly designed to communicate with each other to share information, often at ultra-low bit rates. The schemes create intentional side channels, allowing existing communication infrastructures to adapt to new technologies and address unforeseen challenges.

Recent CTC research has focused on smart devices and the internet of things (IoT) to facilitate simple, yet crucial tasks such as 802.11 Wi-Fi to 802.15.4 Zigbee communication \cite{WEBEE}, improving smart device setup processes \cite{STRAP} or enabling long-range IoT sensor updates from 802.11 devices by modulating the noise floor \cite{OFNP}.

\begin{figure}
    \centering
    \vspace{-1cm}
    \includegraphics[width=1\linewidth]{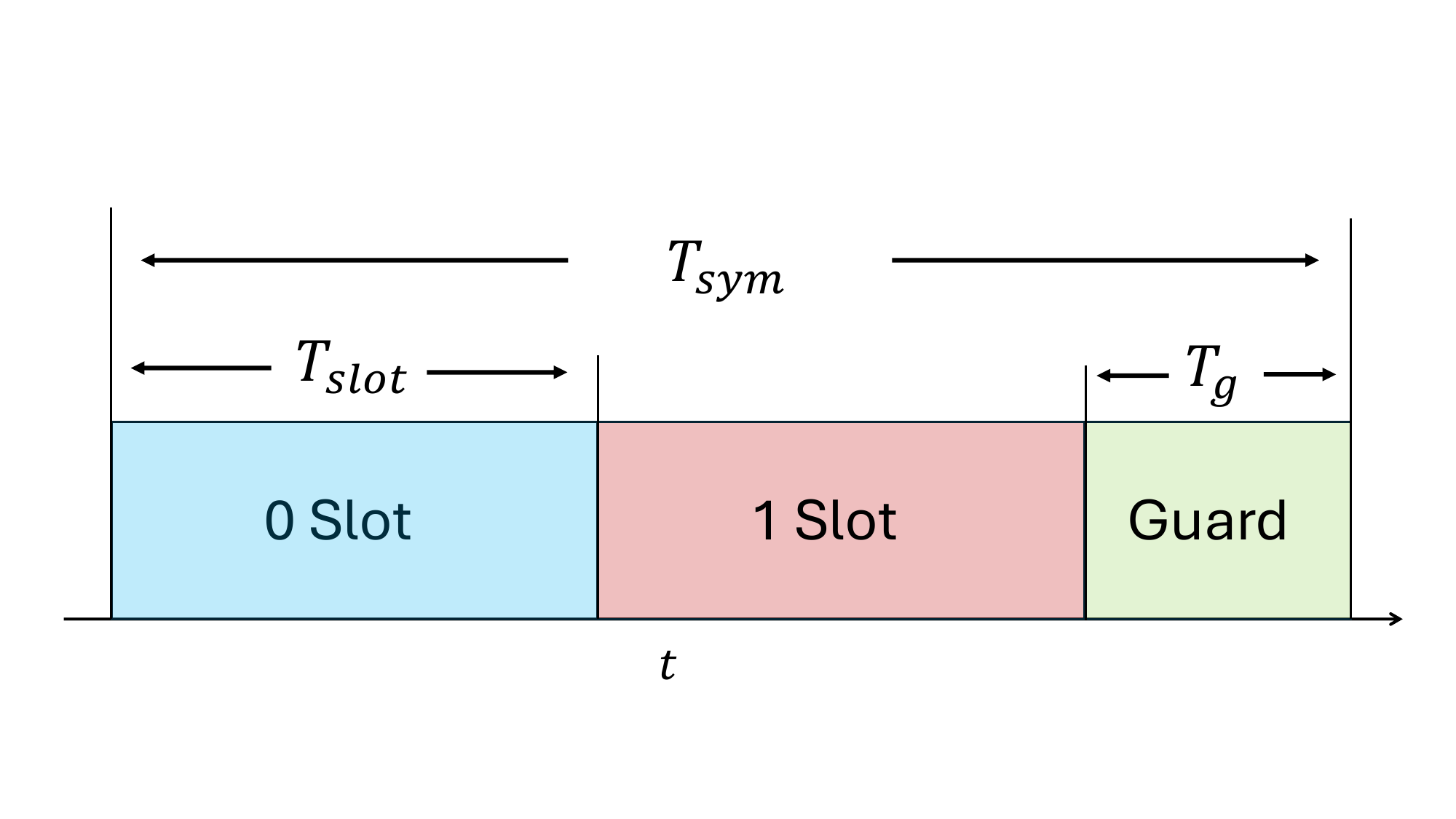}
    \vspace{-1cm}
    \caption{Ghost modulation symbol structure.}
    \label{fig:GM_SlotFigure}
\end{figure}

Ghost Modulation (GM) \cite{GM_1}, a recently proposed CTC protocol, is designed to transmit secondary messages over an existing network for applications like embedding identifying information in satellite internet networks. GM introduces artificial delays into the network stream, treating each packet as a pulse, to communicate the secondary message. This technique is conceptually similar to binary pulse position modulation (PPM). GM is capable of being implemented at any level of the open systems interconnection (OSI) model \cite{OSI_MODEL}, with a particular emphasis on the higher layers, such as the application layer. 

GM presents several intriguing technical challenges. A key challenge is that unlike traditional modulations with strict hardware-based timing control, GM operates without such precision. GM is vulnerable not only to normal channel noise, but the network hops themselves will introduce delays into the packet timing. These delays introduce a second major error mode, distinct from those encountered in typical channels. Even in single-hop networks, packets can be delayed within the operating system or by the network interface controller (NIC). In this work, we present a model for this unusual asymmetric channel and analyze its capacity.

Additionally, while robust receiver time synchronization schemes exist for major PPM variations \cite{PPM_SYNC, PPM_SYNC_AUTOCORR, PPM_SYNC_AND_DETECT, PPM_SYNC_OPPM, PPM_SYNC_OPPM_CODING_THEORY, PPM_SYNC_MPPM, PPM_SYNC_MPPM_SYM}, the network-induced delays that GM encounters differ significantly from traditional timing issues such as clock offsets and jitter. Consequently, existing literature on PPM time synchronization is insufficient to address GM's challenges, necessitating a novel joint acquisition and timing synchronization scheme.

In this work, we apply information-theoretic analysis to the GM signal and channel models and present detection and synchronization algorithms. Section \ref{sec:signal_model} presents and justifies the signal model, the discrete channel model, and channel capacity. Simplified maximum-likelihood decision rules are derived in Section \ref{sec:decision_rules}, forming the basis of a heuristic decision rule. Section \ref{sec:sync} presents a signal acquisition and timing estimation algorithm. Finally, in Section \ref{sec:coding} we apply several well-known, short-block-length, forward error correction (FEC) codes and present the resulting bit error rate (BER) performance.

\section{Signal and Channel Models}\label{sec:signal_model}

\begin{figure}
    \centering
    \includegraphics[width=1\linewidth]{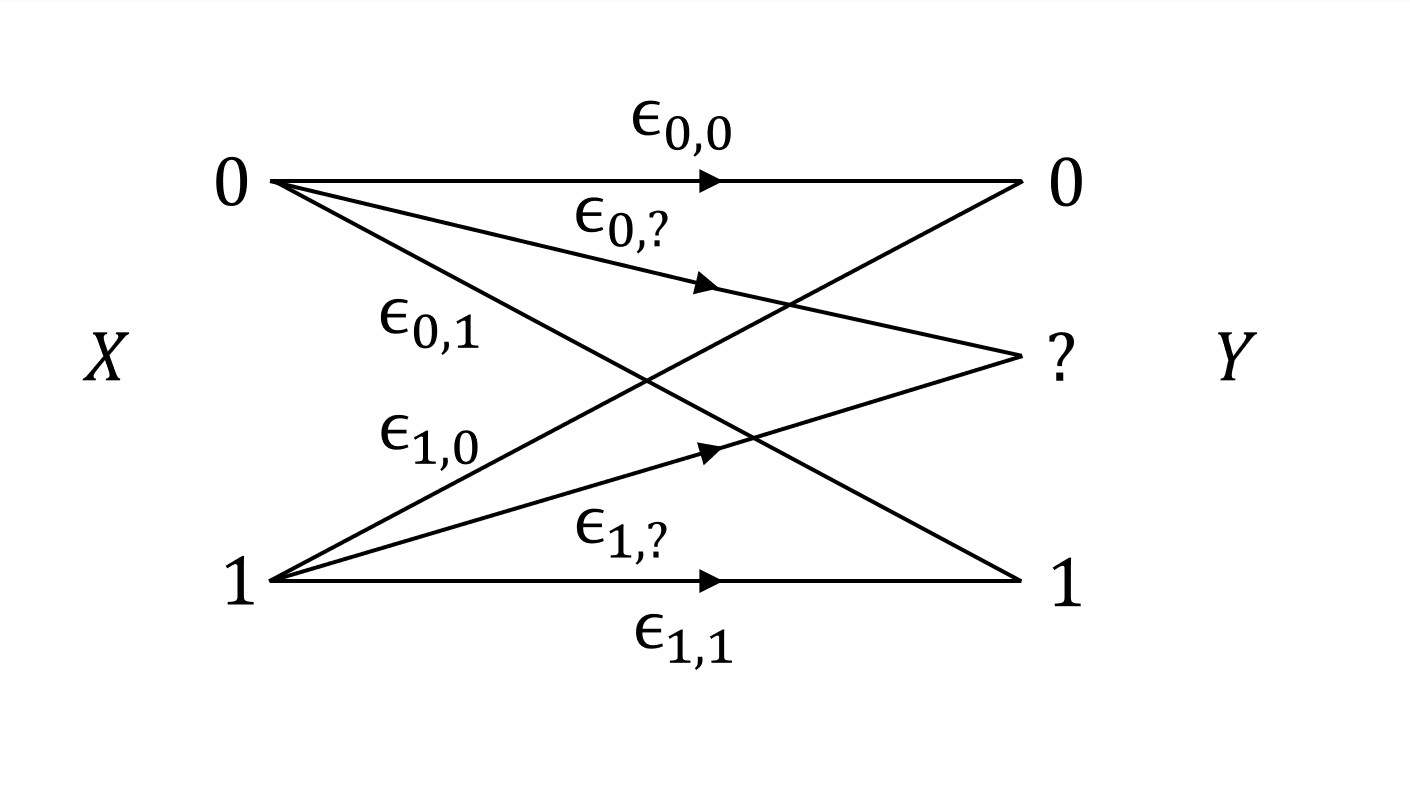}
    \caption{Binary crossover erasure channel with fully-asymmetric transition probabilities}
    \label{fig:channel-model}
\end{figure}

GM operates similarly to PPM, where the pulses are packets in the network. PPM works by transmitting a pulse within one of $M$ time-domain transmission slots of length $T_{\text{slot}}$ that are uniformly spaced within a symbol time, $T_{\text{sym}}$. GM also has guard periods of length $T_{\text{g}}$ after symbol periods, as shown in Fig. \ref{fig:GM_SlotFigure}. 

In the implementation of GM, many practical factors must be considered, such as how to identify network packets acting as pulses, how to maintain covertness, or how to best preserve network throughput. While important, this work instead focuses on the theoretical foundation of GM and, as such, we assume an equivalent signal model for binary GM.

The equivalent continuous-time transmitted signal is modeled as

\begin{equation} \label{eq:signal_model}
    s(t) = \sum_{k=0}^{N-1} \delta(t - kT_{\text{sym}} - \theta_k T_{\text{slot}})
\end{equation}
where $N$ is the total number of symbols, $T_{\text{sym}}$ is the symbol time, $T_{\text{slot}}$ is the slot time for a ``0" or a ``1" time slot, and $\theta_k \in \{0, 1\}$ is the $k$th GM symbol. A guard window occurs after the two pulse windows, of length $T_{\text{g}}$, such that
\begin{equation}
    T_{\text{sym}} = 2 T_{\text{slot}} + T_{\text{g}}.
\end{equation}

The choice of a Dirac delta function to represent each pulse is natural in that (1) packets are not allowed to be directly modified, and (2) all the GM information is contained in the start time of the packet. Additionally, the delta functions allow for easier modeling as they can easily be convolved with envelope functions to better represent the network packets. As GM can be built atop any packetized network, this work does not assume a specific packet type or envelope for generality’s sake. 

A transmitted signal may also be represented in vector form as
\begin{equation}
    \bm{s} =\begin{bmatrix} 
        \theta_0 T_{\text{slot}} \\
        T_{\text{sym}} + \theta_1 T_{\text{slot}} \\
        \vdots \\
        (N-1)T_{\text{sym}} + \theta_{N-1} T_{\text{slot}}
    \end{bmatrix}^\intercal
\end{equation}
where each element is the time stamp of intended transmission. The continuous-time received signal is represented by 
\begin{equation}
    r(t) = \sum_{k=0}^{N-1} z_k \delta(t - kT_{\text{sym}} - \theta_k T_{\text{slot}} - \tau_k)
\end{equation}
where $\tau_k \sim \text{Exp}(\lambda)$ is the network-induced delay, $\frac{1}{\lambda}$ is the mean delay time, $z_k \sim \text{Bernoulli}(1 - \rho)$, and $\rho$ is the packet drop rate of the network. The received vector is 
\begin{equation} \label{eq:vector_signal_model}
    \bm{r} =  (\bm{s} + \bm{\tau}) P^{(N)}
\end{equation}
where $P^{(N)}$ is an $N \times \tilde{N}$ quasi-permutation matrix with
\begin{equation}
    0 \le \tilde N \le N.
\end{equation}
$P^{(N)}$ sorts the values of $(\bm{s} + \bm{\tau})$ to account for their time-ordered reception and for the possibility of dropping packets, and as such, $\bm r$ is a $1 \times \tilde{N}$ vector while $\bm{s}$ is a $1 \times N$ vector.  

\subsection{Discrete  Memoryless Channel Model}
The discrete memoryless channel model in general is a binary crossover erasure channel (BCEC) \footnote{The more common analog of this channel is the binary symmetric erasure channel (BSEC) which features symmetric probabilities of bit-flips and symmetric erasure probabilities.} depicted in Fig. \ref{fig:channel-model}. We denote the transition probabilities from input state $n$ to output state $m$ as $\epsilon_{n,m}$ under the constraints that 
\begin{align}
\epsilon_{0,0} + \epsilon_{0, 1} + \epsilon_{0, ?} = 1 \\
\epsilon_{1,1} + \epsilon_{1, 0} + \epsilon_{1, ?} = 1.
\end{align}
We justify this channel model by looking at some particular errors of note, assuming temporarily that $T_{\text{g}} = T_{\text{slot}}$.

\begin{figure}
    \centering
    \includegraphics[width=1\linewidth]{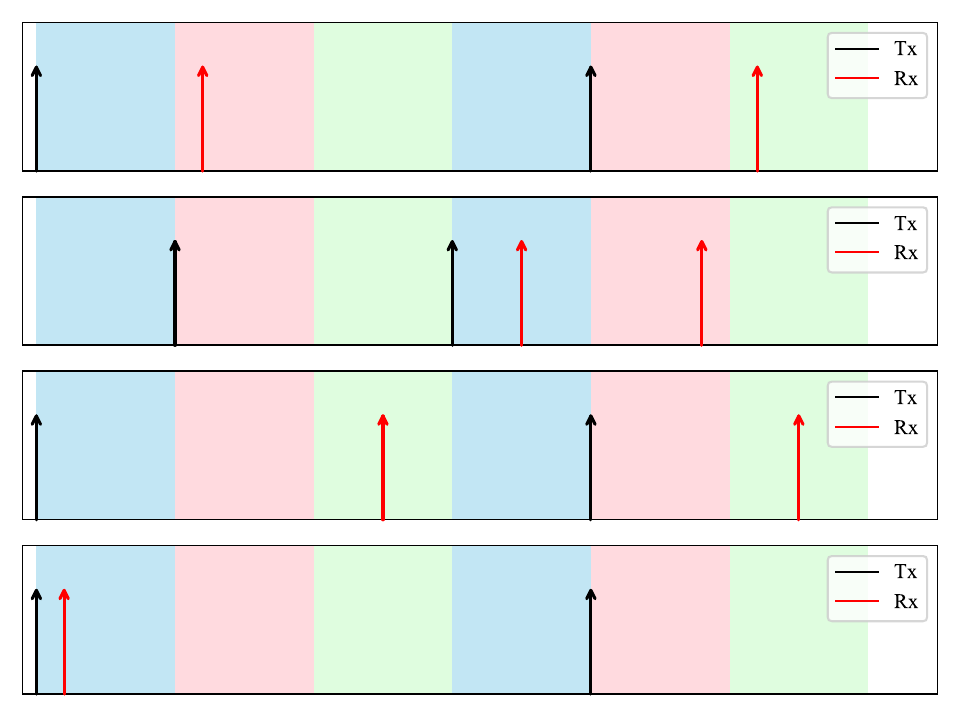}
    \caption{Common errors in Ghost Modulation that demonstrate the asymmetry present in the discrete, memoryless channel model.}
    \label{fig:common_errors}
\end{figure}

Several common errors are depicted in Fig. \ref{fig:common_errors}, where the colored regions align with those used in Fig. \ref{fig:GM_SlotFigure} and the cases are ordered with case one at the top and case four at the bottom. The first case of interest demonstrates that delays of equal length result in asymmetric transitions, i.e., a zero pulse is most likely to be delayed into the one time slot, while a one pulse is most likely to be delayed into the guard period. The second case of interest demonstrates that (1) packets do not contain sequence numbering and thus can't be perfectly time-resolved at the receiver, and (2) it is possible to arrive at a symbol period with no pulses at all or a symbol period with multiple pulses. These problems are further compounded by longer messages, where error-induced ambiguity scales with $N$. Case three is one in which it is unclear whether a one or zero packet was intended, as both transmitted symbols arrive in the guard period. It is, however, immediately clear that it is more likely for a one pulse to arrive in the guard period than a zero pulse, as the exponential probability distribution function (PDF) is monotonically decreasing. Finally, case four presents a plausible reception with packet drops. In this scenario, there is no information present at the receiver about the possible transmitted signal in the second symbol period.

Upon inspection of the possible delay and drop combinations, it is intuitively and empirically clear that (1) the channel has memory between symbol periods and (2) there is asymmetry in the transitions between the input symbols $\{0, 1\}$ and the output symbols $\{0, 1, ?\}$. Thus, we present the equivalent discrete, memoryless channel, depicted in Fig. \ref{fig:channel-model} with transition probabilities that are tied to the delay and packet drop statistics and $T_\text{g}$.

Of particular interest to the field of channel coding is that the BCEC is a relatively unstudied discrete channel model. As most asymmetric channel coding work focuses on the Z-channel and relies heavily on large block length codes such as Polar or Low Density Parity Check (LDPC) codes~\cite{How_to_Achieve_the_Capacity_of_Asymmetric_Channels}, there is little research on how to achieve the capacity of such a channel for the small block length regime that GM operates in.

\subsection{Capacity}

Channel capacity is defined in general as 

\begin{align}
C = \max_{P(x)} I(X; Y) \quad \text{bits per channel use}
\end{align}

where $I(X; Y)$ denotes the mutual information between the discrete random variables $X$ and $Y$, and $C$ represents the supremum of all achievable rates with an arbitrarily low error rate~\cite{shannon1948mathematical, InfoTheoryBook}. We model the binary inputs as $X \sim \text{Bern}(\gamma)$ where $\gamma = P(x = 0)$ and $1 - \gamma = P(x = 1)$. The ternary channel output is $Y$, following the mapping in Fig.~\ref{fig:channel-model}. Noting that $P(x)$ is univariate, we arrive at the expression

\begin{align}
C &= \max_{\gamma} \left(\epsilon_{0,0}\gamma + \epsilon_{1,0}(1-\gamma)\right) \log_2 (\epsilon_{0,0}\gamma + \epsilon_{1,0}(1-\gamma)) \label{eq:capacity} \\
&\quad\quad+ \left( \epsilon_{0,e}\gamma + \epsilon_{1,e}(1-\gamma) \right) \log_2 \left( \epsilon_{0,e}\gamma + \epsilon_{1,e}(1-\gamma) \right) \nonumber \\
&\quad\quad+ \left( \epsilon_{0,1}\gamma + \epsilon_{1,1}(1-\gamma) \right) \log_2 \left( \epsilon_{0,1}\gamma + \epsilon_{1,1}(1-\gamma) \right) \nonumber \\
&\quad\quad+ \gamma (\zeta_1 - \zeta_2) + \zeta_2 \nonumber 
\end{align}
where we define the constants

\begin{align*}
\zeta_1 &\triangleq \epsilon_{0,0} \log_2 \epsilon_{0,0} + \epsilon_{0,?} \log_2 \epsilon_{0,?} + \epsilon_{0,1} \log_2 \epsilon_{0,1} \\
\zeta_2 &\triangleq \epsilon_{1,1} \log_2 \epsilon_{1,1} + \epsilon_{1,?} \log_2 \epsilon_{1,?} + \epsilon_{1,0} \log_2 \epsilon_{1,0}.
\end{align*}
The capacity does not have a closed form expression, but is a trivial optimization problem with regard to $\gamma$. The capacity of this channel is bounded by 

\begin{align}
0 \le C \le 1
\end{align}
and $\gamma$ is conveniently bounded \cite{Two_Results} by

\begin{align}
\frac{1}{e} < \gamma < 1 - \frac{1}{e}.
\end{align}

\section{Decision Rules}\label{sec:decision_rules}

The complex nature of the GM channel is ill-suited for deriving the general maximum-likelihood decision. Instead, we look at two simplified cases that ignore packet drops, then use the maximum-likelihood decision rules from each of those cases as a justification for a heuristic decision rule. The final decision rule utilizes erasures in the output to only address a single symbol period at a time, for simplicity at the receiver.

\subsubsection{Case 1}\label{subsec:case1}

We look first at the case where packets are received according to a Poisson random process, where reordering of the packets does not occur, and no packets are dropped in reception.

Given the maximum-likelihood criteria,
\begin{equation}
    r^{(\text{h}_1)}_i = \argmax_{ \theta_i \in \{1, 0\} } P( r_i | \theta_i )
\end{equation}
we say that the optimal decision rule is
\begin{equation} \label{eq:simple_decision_rule}
    r_i^{(\text{h}_1)} = 
    \begin{cases}
        0 & i T_{\text{sym}} \le r_i \le i T_{\text{sym}} + T_{\text{slot}} \\
        1 & \text{Otherwise}
    \end{cases}
\end{equation}
because $\tau_i \sim \text{exp}(\lambda)$ we can say
\begin{align}
P(r_i | 0) > P(r_i | 1) \quad&\text{for } (i-1) T_{\text{sym}} \le r_i < (i-1) T_{\text{sym}} +  T_{\text{win}} \\
P(r_i | 0) < P(r_i | 1) \quad&\text{for } (i-1) T_{\text{sym}} + T_{\text{win}} \le r_1 .
\end{align}
The received vector $\mathbf{r}$ is converted into the binary vector
\begin{equation}
    \mathbf{r}^{(\text{h}_1)} = 
    \begin{bmatrix}
        r^{(\text{h}_1)}_0 & r^{(\text{h}_1)}_1 & r^{(\text{h}_1)}_2 & \cdots & r^{(\text{h}_1)}_{n-1}
    \end{bmatrix}
\end{equation}
using \eqref{eq:simple_decision_rule}.

\subsubsection{Case 2} Second, we look at the case where reordering can happen (under the assumption that the mean delay is “small” i.e., with a mean delay far less than $T_{\text{slot}}$), where we still do not consider packet drops.


Starting from the maximum likelihood position, we must consider the whole received vector at a time, because of the time-dependence inherent to reordering. Thus, we have,
\begin{equation}
    \textbf{r}^{(\text{h}_2)} = \argmax_{\bm \theta \in \{0, 1\}^{(N)}} P(\textbf{r} | \bm{\theta}).
\end{equation}
Using the law of total probability, we can state that
\begin{equation}
    P(\textbf{r} | \bm{\theta}) = \sum_{j=0}^{N-1} \sum_{i=0}^{N-1} P(\textbf{r} | \bm{\theta}, r_i \rightleftharpoons \theta_j) P(r_i \rightleftharpoons \theta_j)
\end{equation}
where $\rightleftharpoons$ denotes correspondence between $\theta_j$ and $r_i$, i.e., that channel input symbol $c_j$ produced output symbol $r_i$. Using the small delay assumption and the knowledge that some received symbols $r_i$ cannot correspond to certain input symbols (i.e., if they are smaller than $(j-1)T_{\text{sym}}$ of the corresponding $\theta_j$), we can then say that
\begin{equation}
    P(r_i \rightleftharpoons \theta_j) \approx 0 \quad i \ne j
\end{equation}
and 
\begin{equation}
    P(\textbf{r} | \bm \theta) \approx \sum_{i=0}^{N-1} P(\textbf{r} | \bm \theta, r_i \rightleftharpoons \theta_i) P(r_i \rightleftharpoons \theta_i)
\end{equation}
which then reduces down to \eqref{eq:simple_decision_rule}.

\subsection{Final Decision Rule}

With the introduction of packet drops into our analysis, we now have too much information lost about whether an error occurs due to a packet being delayed into another symbol period or from missing packets altogether. To ensure a simple but likely non-optimal decision rule, we choose to look at each symbol period independently, and make decisions only on the information present within each symbol period. Thus, we arrive at the final decision rule as 
\begin{equation} \label{eq:final_decision_rule}
    r_{i}^{(\text{e})} = 
    \begin{cases}
    0 & h(\bm r, i) \text{ and } \exists \: j \: | \: i T_{\text{sym}} \le r_j < i T_{\text{sym}} + T_{\text{slot}} \\
    1 & h(\bm r, i) \text{ and } \\
    &\exists \: j \: | \: i T_{\text{sym}} + T_{\text{slot}} \le r_j < i T_{\text{sym}} + 2T_{\text{slot}} \\
    ? & \text{otherwise}
    \end{cases}
\end{equation}
where we define the indicator function
\begin{equation}
    h(\bm r, i) = \begin{cases}
        1 &  |\{ j \: | \: i T_{\text{sym}} \le r_j < (i+1) T_{\text{sym}} \}| = 1\\
        0 & \text{otherwise}
    \end{cases}
\end{equation}
and $|\cdot|$ in this context refers to the cardinality of a set. Notionally, we change the superscript $^{(\text{h})}$ to $^{(\text{e})}$ to emphasize the change from hard binary decisions to hard ternary decisions. The final output vector is
\begin{equation}
\bm r^{(\text{e})} = \begin{bmatrix}
        r^{(\text{e})}_0 & r^{(\text{e})}_1 & r^{(\text{e})}_2 & \cdots & r^{(\text{e})}_{N-1}
    \end{bmatrix},
\end{equation}
a strictly $1 \times N$ row vector. As \eqref{eq:final_decision_rule} looks at each symbol period independently, the issue evident from \eqref{eq:vector_signal_model}, where there is no guarantee that the number of elements in $\bm r$ is equal to the number of elements in $\bm s$, is overcome. 

\section{Signal Acquisition and Synchronization}\label{sec:sync}

As discussed in the introduction, there already exist many time synchronization schemes for PPM. Many start from the assumption of slot synchronization \cite{PPM_SYNC_AND_DETECT}, which may be an adequate assumption for optical transmission networks or networks with alternate timing synchronization mechanisms. Recall that the primary differences between the GM signal model and the traditional PPM signal model are (1) the random transmission delays that occur in the process of delaying a packet and (2) the assumption of continued channel use (i.e., GM is built atop an existing network that will still operate in the absence of a secondary message). Thus, we encounter the issue of how to detect the start of a GM signal and perform time synchronization to align the time slots in the presence of a continually used network channel and random delays.

To our knowledge, there is no existing literature to rely on as a basis for solving this problem. The closest analogs to this issue are the classical timing synchronization problem in traditional communications theory such as QAM and QPSK and time dispersive channels as seen in high-mobility scenarios. Neither of these provide assistance in this case of timing noise. Another interesting area of research, that appears to be similar, is the problem of synchronizing computer clocks across networks. In this scenario, many packets are sent from a server containing the true time (usually a server attached to an atomic clock) and the average delay is estimated, leading to an accurate time estimate. The most famous of these algorithms is the Network Timing Protocol (NTP) \cite{NTP_IEEE}. This fails to help in the GM case of timing noise, because the packets cannot be altered themselves. The delays mean that information is lost about the true ordering of the packets, and no assumptions are made about the content of the packets themselves, and thus the GM system cannot implement or adapt a time synchronization scheme from the wealth of research available from the field of networking timing or PPM.

Traditional techniques for signal acquisition in digital communication are usually based on time-domain, deterministic autocorrelation \cite{RICE_TEXTBOOK, PROAKIS, SCHMIDL}. These techniques work well in common cases and are usually modified for situations such as in time dispersive channels, e.g., the high-velocity channel. These techniques do not account for the random delay that is found in the GM channel model. The clearest analog for random delays in traditional communications literature is that of clock jitter, where irregularities and imperfections in the hardware clock results in slightly irregular sampling, making symbol decisions difficult as information leaks between symbol periods. While similar, it cannot be solved in the same ways because of the magnitude of the delays and because the source of the timing noise is very different in nature than that of clock jitter.

\subsection{Signal Acquisition and Time Synchronization Algorithm}

We propose an ad-hoc detection and time-synchronization method inspired by matched filtering and maximum-likelihood estimation. Given an arbitrary function $f(t)$, we define the discrete version of the function according to the indicator function
\begin{equation} \label{eq:bin_function}
f[n] = \begin{cases}
1 & \exists \:\: q \: | \: f(q) \ne 0  \text{ and } tT_{\text{bin}} \le q < (t + 1) T_{\text{bin}} \\
0 & \text{otherwise}
\end{cases}
\end{equation}
resulting in a discrete-time, sampled version of the function, where $t = n T_{\text{bin}}$ for $n \in \mathbb Z$. $T_{\text{bin}}$ is some positive value such that the number of bins per symbols $N_{\text{bins}} = T_{\text{sym}} / T_{\text{bin}}$ is a positive integer. This form of discretization is commonly called binning, rather than typical sampling. This is a direct result of using $\delta(t)$ in \eqref{eq:signal_model}.

Next, given some vector $\bm{\theta}_P$ of length $N_P$ that represents the known preamble sequence, the known impulse train of the preamble is
\begin{equation} \label{eq:template_preamble}
    S_P(t) = \sum_{k=0}^{N_P-1} \delta(t - kT_{\text{sym}} - \theta_{P,k} T_{\text{slot}})
\end{equation}
where we then use \eqref{eq:bin_function} to create the discrete-time version $S_P[n]$. In a typical system, this might be used in a correlation-based detection metric, however with the timing error and impulses, it will certainly fail. Instead, we can spread the impulses out, and in essence, weight the received pulses by how close they are to the known intended impulse positions. We do this by creating the new correlation signal as
\begin{equation} \label{eq:mf_conv}
    Q [n] = S_P[n] * f^{'}_{\tau} [n]
\end{equation}
where $*$ represents discrete convolution, and 
\begin{equation}
f^{'}_{\tau} (t) = 
    \begin{cases}
        \lambda^{'} e^{-t  \lambda^{'}} & 0 \le t < T_{\text{win}} + T_{\text{g}} \\
        0 & \text{otherwise}
    \end{cases}
\end{equation}
is the truncated probability distribution function for $\tau_k$, the random delay at time $k$. We choose to use $\lambda^{'}$ as a parameter of this detector instead of $\lambda$ as $\lambda$ is unlikely to be known or have been estimated at signal acquisition. Using the sifting property of discrete convolution, we can write \eqref{eq:mf_conv} as 
\begin{equation}
    Q[n] = \sum_{k=0}^{N_P-1} f^{'}_{\tau}\left(n T_\text{bin} - kT_{\text{sym}} - \theta_{P,k} \right)
\end{equation}
which is correlated with the incoming signal $s[n]$ according to
\begin{equation}\label{eq:detection_metric}
    M[n] = \sum_{m=0} s[m] Q[n + m]
\end{equation}
which is the detection metric. Observe that the maximum value of $M[n]$ is constrained. In the no delay case, the maximum output of the signal is $N_P \lambda^{'}$, where all the impulses align exactly with values that correspond to $f^{'}(0)$. This is very convenient for setting a detection threshold as there is no channel gain term present. When packets drop, they only detract from the metric and rarely do valid network packets burst into existence. Thus, if a peak is sufficiently close to $N_P \lambda^{'}$, it indicates the presence of the preamble and consequently provides timing as to which set of time bins align with which modulation slots.

In the case where no packets were dropped, and no packets were delayed past their respective symbol period, the magnitude of the detected point provides an ideal time to estimate the delay parameter $\lambda$. Given the known vector of known preamble timestamps $\bm S_P$ derived from \eqref{eq:template_preamble} and the set of timestamps that were found to correspond to the received preamble $\bm r_P$, we can then measure the distances between the intended start times and the actual received timestamps. The estimate of the mean delay is
\begin{equation}
\frac{1}{\hat \lambda} = \frac{1}{N_P} \sum_{i=0}^{N_P} s_i - (r_i - \hat t_p)
\end{equation}
where $\hat t_p = \hat n_p T_{\text{bin}}$ is the estimated start time from \eqref{eq:detection_metric} ($\hat n_p$ is discussed below).
\subsection{Detection Algorithm Performance}

\begin{figure}
    \centering
    \includegraphics[width=1\linewidth]{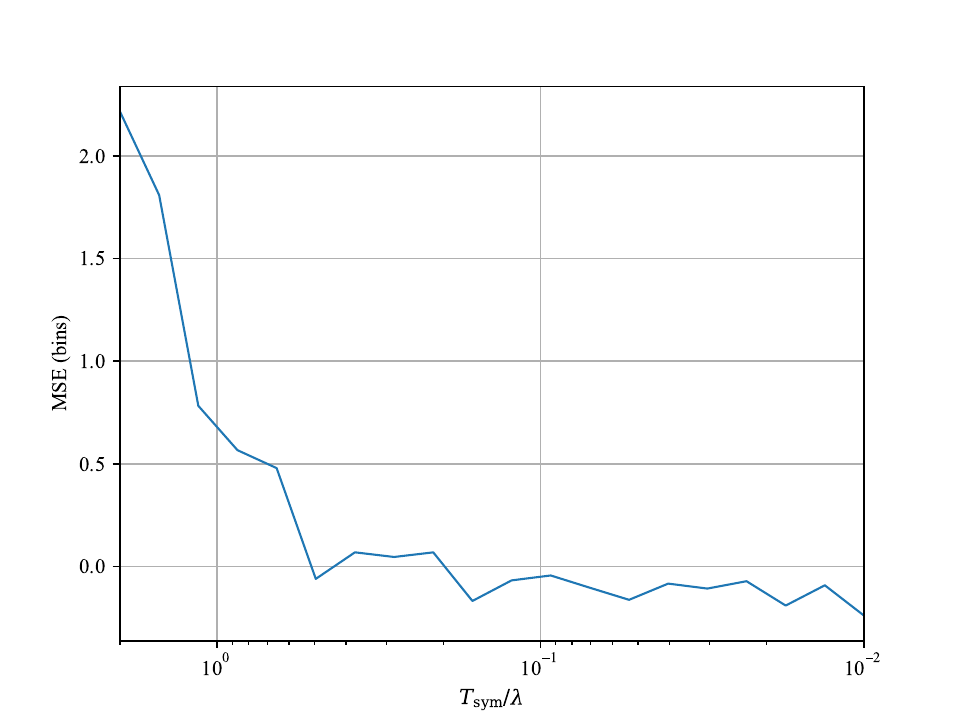}
    \caption{Mean squared error of the proposed detection algorithm in number of time bins away from the true signal start location for varying levels of normalized mean delay.}
    \label{fig:detection_MSE}
\end{figure}

To test the algorithm, we generated a constant stream of GM modulated data and embedded a randomly generated preamble within the stream for each trial, then each pulse was delayed and some were randomly dropped from the signal. Exact GM parameters are shown in Table~\ref{tab:detection_sim_params}. The detection algorithm was then run on the generated signal and results are shown in Fig. \ref{fig:detection_MSE} where the max over the range was determined to be the estimated preamble start $\hat{n}_{P}$. where we define mean squared error (MSE) as
\begin{equation}
MSE = \sum_{k=0}^{K} | \hat{n}_{P} - n_{P} |^2
\end{equation}
for a true preamble start at $n_{P}$. Observe that at very low mean delays, the algorithm under-performs, hitting an error floor (in this case a very small error floor). This occurs due to the binning effect, wherein the precision contained within $\bm r$ is lost. The obvious trade-off then is that more bins are required to utilize the benefits of a good channel, which means more computations and more memory. Another key assumption that became clear in simulation was that, while modeled as impulses, there is in fact a minimum spacing between packets (if GM is implemented in an 802.11 style network) due to the minimum length of a 802.11 packet resulting from the network overhead in the 802.11 preamble and header structure. This underlying assumption requires protections against multiple impulses arriving in the same time bin during simulation.

Future works could explore the exact nature of probability of false alarm and probability of missed detection given the ability to correctly decode a GM packet and the tolerance of slot misalignment by one or more time bins.

\begin{table}[htbp]
    \caption{Detection Algorithm Testing Parameters}
    \label{tab:detection_sim_params}
    \begin{center}
    \begin{tabular}{ |c|c|  }
        \hline
        \multicolumn{2}{|c|}{Simulation Parameters} \\
        \hline
        $T_{\text{sym}}$&0.04\\
        $T_{\text{slot}}$&0.0175\\
        $T_{\text{guard}}$&0.005\\
        $N_{P}$&30\\
        $N_{\text{slots}}$& 50\\
        $\lambda^{'}$&$\lambda$\\
        $\rho$&2\%\\
        \hline
    \end{tabular}
    \vspace{0.5cm}
    \end{center}
\end{table}

\section{Coding Results}\label{sec:coding}

\begin{figure}
    \centering
    \includegraphics[width=1\linewidth]{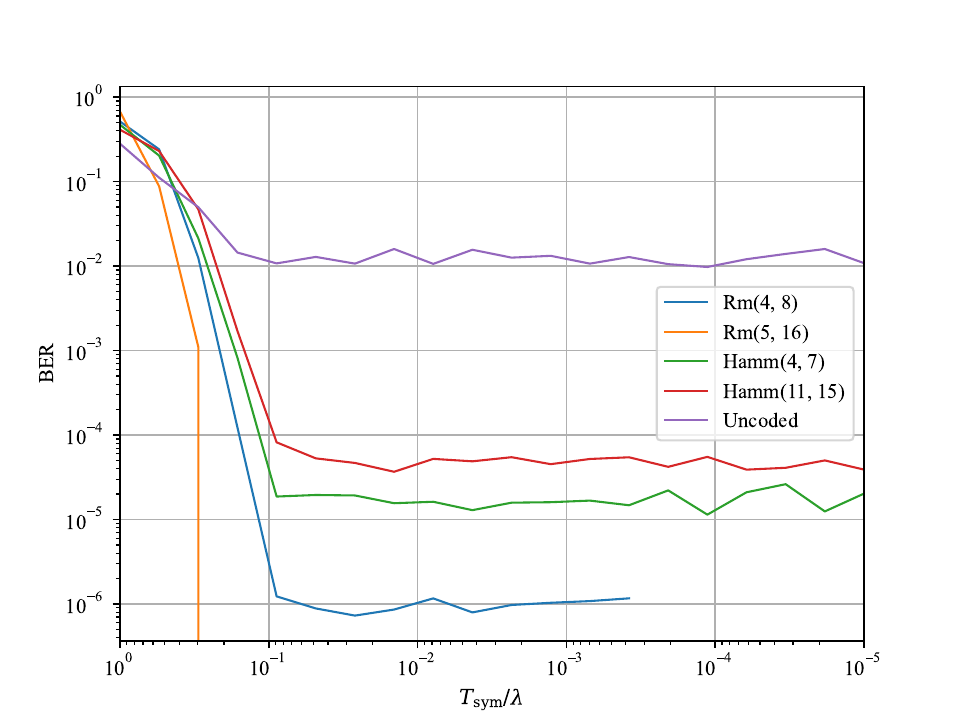}
    \caption{Coding results for GM and the BCEC channel. Here, Rm and Hamm refer to Reed-Muller and Hamming codes respectively. The first argument is the dimension, $k$, and the second is the block length, $n$.}
    \label{fig:gm_coding results}
\end{figure}

Given that capacity achieving codes exist for the binary symmetric and binary erasure channels, perhaps eventually generalized capacity achieving codes for the BCEC will materialize in the future. In the meantime, we applied several well known short block length linear codes to GM to determine their performance. The rate-adjusted results for this are shown below over a range of normalized mean delays in Fig \ref{fig:gm_coding results}. 

Simulation parameters are the same as those used in Section \ref{sec:sync}, shown in Table \ref{tab:detection_sim_params}. The packet drop rate, $\rho$, was fixed at 0.02 or 2\% accounting for the perceived limit on performance. Due to the fixed packet drop rate and the computations required to get error rates sufficiently low, BER variance appears to be higher than it might actually be in practice.


\section{Conclusion}\label{sec:conclusion}

In this work, we presented and justified the signal model and discrete channel model for the recently proposed Ghost Modulation. The channel capacity was presented along with an examination of the dual error modes. We derived two simplified maximum-likelihood decision rules, forming the basis for a proposed heuristic decision rule. As the existing literature on time synchronization is insufficient for detecting GM preambles, we proposed a joint signal acquisition and time synchronization algorithm and justified the scheme with simulation results. Finally, we applied well-known error correction codes and presented their performance.

\bibliographystyle{IEEEtran}
\bibliography{refs}

\end{document}